\documentclass[sigchi]{acmart}

\AtBeginDocument{%
	\providecommand\BibTeX{{%
			\normalfont B\kern-0.5em{\scshape i\kern-0.25em b}\kern-0.8em\TeX}}}

\usepackage{subfigure}
\usepackage{mathrsfs}
\usepackage[linesnumbered, boxed, ruled, commentsnumbered]{algorithm2e}

\renewcommand\footnotetextcopyrightpermission[1]{}
\begin{document}
	

\title{FraudJudger: Real-World Data Oriented Fraud Detection on Digital Payment Platforms}
 
\author{Ruoyu Deng}
\affiliation{%
	\institution{Shanghai Jiao Tong University}
	\state{China} 
}
\email{dengruoyu@sjtu.edu.cn}

\author{Na Ruan}
\affiliation{%
	\institution{Shanghai Jiao Tong University}
	\state{China} 
}
\email{naruan@sjtu.edu.cn}
 
	\begin{abstract}
	Automated fraud behaviors detection on electronic payment platforms is a tough problem. Fraud users often exploit the vulnerability of payment platforms and the carelessness of users to defraud money, steal passwords, do money laundering, etc, which causes enormous losses to digital payment platforms and users. There are many challenges for fraud detection in practice. Traditional fraud detection methods require a large-scale manually labeled dataset, which is hard to obtain in reality. Manually labeled data cost tremendous human efforts. Besides, the continuous and rapid evolution of fraud users makes it hard to find new fraud patterns based on existing detection rules. In our work, we propose a real-world data oriented detection paradigm which can detect fraud users and upgrade its detection ability automatically. Based on the new paradigm, we design a novel fraud detection model, FraudJudger, to analyze users behaviors on digital payment platforms and detect fraud users with fewer labeled data in training. FraudJudger can learn the latent representations of users from unlabeled data with the help of Adversarial Autoencoder (AAE). Furthermore, FraudJudger can find new fraud patterns from unknown users by cluster analysis. Our experiment is based on a real-world electronic payment dataset. Comparing with other well-known fraud detection methods, FraudJudger can achieve better detection performance with only 10\% labeled data.
	\end{abstract}
	
	
	
	

	\maketitle
	
	\section{Introduction}
	Digital payment refers to transactions that consumers pay for products or services on the Internet. With the explosive growth of electronic commerce, more and more people choose to purchase on the Internet. Different from traditional face-to-face payments, digital transactions are ensured by a third-party digital payment platform. The security of the third-party platform is the primary concern. The digital payment brings huge convenience to people's daily life, but it is vulnerable to cybercrime attacks \cite{west2016intelligent} \cite{yao2017automated}. There are many kinds of fraud behaviors. For example, fraudsters may pretend to be a staff in a digital payment platform and communicate with normal users to steal valuable information. Some fraudsters will use fake identities to transact in these platforms. An estimated 73\% of enterprises report some form of suspicious activity that puts around \$7.6 of every \$100 transacted at risk \cite{DigitalPaymentsFraud}. Those frauds cause tremendous damage to companies and consumers. 
	
    \textbf{Challenges.} Automatic detection for fraud payments is a hot topic in companies and researchers. Many researchers focus on understanding fraud users' behavior patterns. It is believed that fraud users have different habits comparing with benign users. The first challenge is how to find useful features to distinguish fraud users with benign users. Sun et al. \cite{sun2018fraudvis} use the clickstream to understand user's behavior and intentions. Some other features like transaction records \cite{zheng2018generative}, time patterns  \cite{kc2016temporal} and illicit address information  \cite{lee2019cybercriminal}, etc, are also proved useful in fraud detection. Fraud users always have social connections. Some researchers focus on analyzing user's social networks to find suspicious behaviors \cite{beutel2013copycatch}  \cite{Varol2017OnlineHI} by graph models. They believe fraud users have some common group behaviors.

	The limitation of the above methods is that it is hard to manually find appropriate features to detect frauds. In traditional fraud detection methods, researchers should try many features until the powerful features are found, and these features may be partial in practice. With the help of deep learning, we can automatically learn the best "feature". Autoencoder \cite{liou2014autoencoder} is an unsupervised model to learn efficient data codings. It can get rid of "noise" features and only leave essential features. Origin features are encoded to latent representations by autoencoder. Makhzani et al.  \cite{makhzani2015adversarial} combine autoencoder and generative adversarial network (GAN) \cite{goodfellow2014generative}, and propose a novel model called "adversarial autoencoder (AAE)". AAE can generate data's latent representations matching the aggregated posterior in an adversarial way. 
	
	Another challenge is lacking sufficient and convincing manually labeled data in the real world. Manually labeled data are always hard to obtain in reality. It costs a vast human resource to identify fraud users manually \cite{viswanath2014towards}. Besides, the human's judgment is sometimes subjective, and it is difficult to detect cunning cheaters. Some researchers use unsupervised learning or semi-supervised learning models to detect frauds \cite{de2018tax}. However, for unsupervised learning, it is hard to set targets and evaluate the performance in training models.
	
	New deceptive patterns of fraud users appear everyday. With the development of detection methods, fraud users also evolve quickly to anti-detection. It is impossible to find all detection rules manually. Labeled data are based on historical experience, and it is hard to find unseen fraud patterns by using labeled data. Since unsupervised learning has no past knowledge \cite{carcillo2019combining}, we can use unsupervised learning methods to find new fraud patterns.

	In our work, we aim at overcoming these real-world challenges in fraud detection. We aim at analyzing users' behaviors and detecting fraud users with a small ratio of labeled data. Furthermore, we want to play an active role in the competition between fraud detection and anti-detection. We aim at detecting potential fraud users who cannot be detected by existing detection knowledge.
	
	\textbf{FraudJudger.} We propose a fraud detection model named FraudJudger to detect digital payment frauds automatically. Our detection model contains three steps. First, we merge users' operation and transaction data. Then merged features are converted to latent representations by adversarial autoencoder. We can use these latent representations to classify users. It is a semi-supervised learning process, which means we only need a few labeled data to train the classification model. Finally, new fraud patterns can be detected by cluster analysis. 
	
	\textbf{Contributions.} In summary, our work makes the following main contributions:
	\begin{enumerate}
		\item We design a novel automated fraud detection paradigm for real-world application. Based on the paradigm, we propose a digital payment fraud detection model FraudJudger to overcome the shortcomings of real-world data. Our model requires fewer labeled data and can learn efficient latent features of users.
		\item Our experiment is based on real-world data. The experiment result shows that our detection model achieves better detection performance with only 10\% labeled data compared with other well-known supervised methods. We propose a new measurement \textit{Cluster Recall} to evaluate cluster results, and our model outperforms others. 
		\item Our model can discover potential fraud users who can not be detected by existing detection rules, and analyzing behaviors of these potential fraud users can help companies build a safer payment environment.
	\end{enumerate}
	
	\textbf{Roadmaps.} The remainder of the paper is organized as follows. In Section~\ref{section:Related Work}, we present related work. Our improved detection paradigm is provided in Section~\ref{section:Detection Paradigm}. Section~\ref{section:Detection Model} presents the details of our detection model FraudJudger. Our experiment is shown in Section~\ref{section:Experiment}. Finally, we conclude our research in Section~\ref{section:Conclusion}.

    \section{Preliminaries}
	\label{section:Related Work}
	\subsection{Digital Payment Fraud Detection}
	Recently, fraud detection on digital payment platforms becomes a hot issue in the finance industry, government, and researchers. There is currently no sophisticated monitoring system to solve such problems since the digital payment platforms have suddenly emerged in recent years. Researchers often use financial fraud detection methods to deal with this problem. The types of financial fraud including credit card fraud \cite{fu2016credit}, telecommunications fraud \cite{kabari2016telecommunications}, insurance fraud \cite{xu2011random} and many researchers regard these detection problems as a binary classification problem. Traditional detection method uses rule-based systems \cite{bahnsen2016feature} to detect the abnormal behavior, which is eliminated by the industry environment where financial fraud is becoming more diverse and updated quickly. With the gradual maturity of machine learning and data mining technologies, some artificial intelligence models have gradually been applied to the field of fraud detection. The models most favored by researchers are Naive Bayes (NB), Support Vector Machines (SVM), Decision Tree, etc. However, these models have a common disadvantage that it is easy to overfit the training data for them. In order to overcome this problem, some models based on bagging ensemble classifier \cite{zareapoor2015application} and anomaly detection \cite{ahmedsurvey} are used in fraud detection. Besides, there are also researchers who use an entity relationship network \cite{van2016gotcha} to infer possible fraudulent activity. In recent years, more and more deep learning models are proposed. Generate adversarial network (GAN) \cite{goodfellow2014generative} is proposed to generate adversarial samples and simulate the data distribution to improve the classification accuracy, and new deep learning methods are applied in this field. Zheng et al. \cite{zheng2018generative} use a GAN based on a deep denoising autoencoder architecture to detect telecom fraud.
	
	Many researchers focus on the imbalanced data problem. In the real world, fraud users account for only a small portion, which will lower the model's performance. Traditional solutions are oversampling minority class \cite{chawla2002smote}. It does not fundamentally solve this problem. Zhang et al. \cite{zhang2018novel} construct a clustering tree to consider imbalanced data distribution. Li et al. \cite{li2014spotting} propose a Positive Unlabeled
	Learning (PU-Learning) model that can improve the performance by utilizing positive labeled data and unlabeled data in detecting deceptive opinions.
	
	Some researchers choose unsupervised learning and semi-super-\\vised learning due to lack of enough labeled data in the real-world application. Unsupervised learning methods require no prior knowledge of users' labels. It can learn data distributions and have potential in finding new fraud users. Zaslavsky et al. \cite{zaslavsky2006credit} use Self-Organizing Maps (SOM) to analyze transactions in payment platforms to detect credit card frauds. Roux et al. \cite{vincent2010stacked} proposed a cluster detection based method to detect tax fraud without requiring historic labeled data. 
	
	In our work, we use semi-supervised learning to detect fraud users, and an unsupervised method is applied in analyzing fraud users patterns and finding potential fraud users. 
	\subsection{Adversarial Autoencoder}
	Adversarial Autoencoder (AAE) is proposed by Makhzani et al \cite{makhzani2015adversarial}. AAE is a combination of autoencoder (AE) and generative adversarial network (GAN). Like GAN, AAE has a discriminator part and a generative part. The encoder part of an autoencoder can be regarded as the generative part of GAN. The encoder can encode inputs to latent vectors. The mission of AAE's discriminator is discriminating whether an input latent vector is fake or real. The discriminator and generator are trained in an adversarial way. AAE is trained in an unsupervised way, and it can be used in semi-supervised classification.
	
	\subsubsection{Autoencoder}
	AE is a feedforward neural network. The information in its hidden layer is called latent variable $z$, which learns latent representations or latent vector of inputs. Recently, AE and its variants like sparse autoencoder (SAE)  \cite{ng2011sparse}, stacked autoencoder \cite{xu2014stacked} have been widely used in deep learning. Basic autoencoder consists of two parts: the encoder part and the decoder part. Encoders and decoders consist of two or more layers of fully connected layers. The encoding process is mapping the original data feature $x$ to the low-dimensional hidden layer $z$. 
	
	\begin{equation}
	z = p(x)
	\end{equation}
	
	The decoding process is reconstructing the latent variable $z$ to the output layer $x'$ whose dimension is the same as $x$.
	\begin{equation}
	x' = q(z)
	\end{equation}
	
	Finally, it optimizes its own parameters according to the Mean Square Error loss (MSE-loss) of $x$ and $x'$.
	\begin{equation}
	AE_{loss} = \frac{1}{n}\sum_{i = 1}^{n}{(x - x')}^2
	\end{equation}
	
	\subsubsection{Generative adversarial autoencoder}
	GAN is first proposed by Ian Goodfellow et al. \cite{goodfellow2014generative}  as a new generation model. Once proposed, it becomes one of the most popular models in the deep learning field. The model mainly consists of two parts, generator $G$ and discriminator $D$. The model can learn the prior distribution $p(x)$ of training data and then generate data similar to this distribution. The discriminator model is a binary-classifier that is used to distinguish whether the sample is a newly generated sample or the real sample. GAN is proposed based on game theory, and its training process is a process of the mutual game. In the beginning, the generator generates some bad samples from random noise, which is easily recognized by the discriminator, and then the generator can learn how to generate some samples that make the discriminator difficult to discriminate or even misjudge. In each training round, GAN will update itself through the process of loss back-propagation. After multiple rounds of the game, the fitting state is finally reached, that is, the generator can generate samples that hard to be distinguished by the discriminator. Here is the loss function of GAN:
	\begin{equation}
	\mathop{min}\limits_{G}\mathop{max}\limits_{D}\mathbb{E}_{x\sim {p_{data}}}[\log{D(x)}] + \mathbb{E}_{z\sim p(z)}\log{(1 - D(G(z)))}
	\end{equation}
	
	where $G$ represents the generator, $D$ represents the discriminator, and $D(x)$ represents a neural network that computes the probabilities that $x$ is real-world samples rather than the generator. $p(z)$ represents the distribution of the noise samples $z$, and $D(G(z))$ represents a neural network that computes the probabilities that a sample is generated by the generator.
	
	\subsubsection{Comparing AAE to GAN}
	Adversarial autoencoder and generative adversarial network have many similarities in unsupervised learning. Both of them can learn data distributions. However, when dealing with discrete data, it will be hard to backpropagate gradient by GAN \cite{yu2017seqgan}. Many discrete features are encoded by one-hot encoding methods. The discriminator of GAN may play tricks that it only needs to discriminate whether the inputs are encoding in the one-hot format. AAE learns latent representations of input features and encodes them as continuous data. The discriminator part of AAE focuses on discriminating latent representations rather than initial inputs so that the original data type of input features does not matter. So AAE has advantages in dealing with discrete features comparing with GANs. In digital payment platforms, many features are discrete, which need to be encoded in one-hot format. It is more reasonable to choose AAE.

	\section{Fraud Detection Paradigm}
	\label{section:Detection Paradigm}
	In this section, we present the traditional fraud detection paradigm on the electronic payment platform first, and then we show our improved paradigm.
	
	\subsection{Traditional Fraud Detection Paradigm}
\begin{figure}[htb]
		\centering
		\subfigure[Traditional fraud detection paradigm ]{
			\centering
			\includegraphics[width=0.42\textwidth]{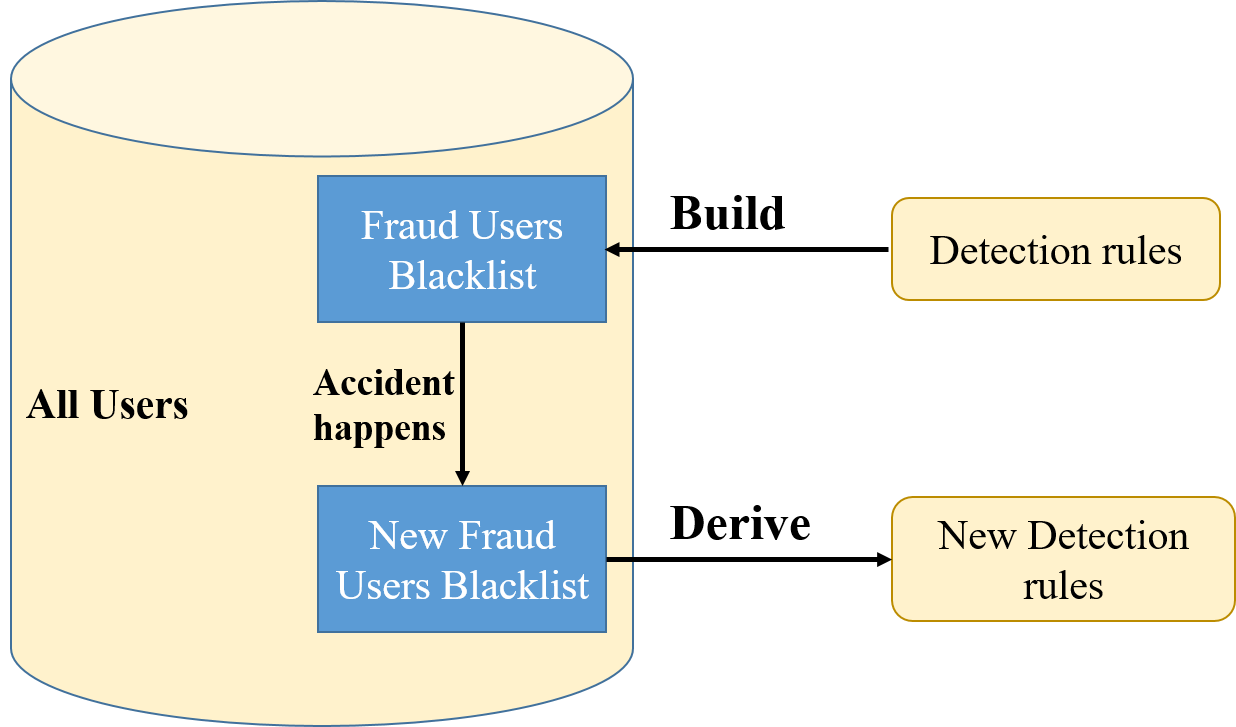}
			\label{figure:paradigm1}
		} 
		
		\subfigure[Improved fraud detection paradigm ]{
			\centering
			\includegraphics[width=0.42\textwidth]{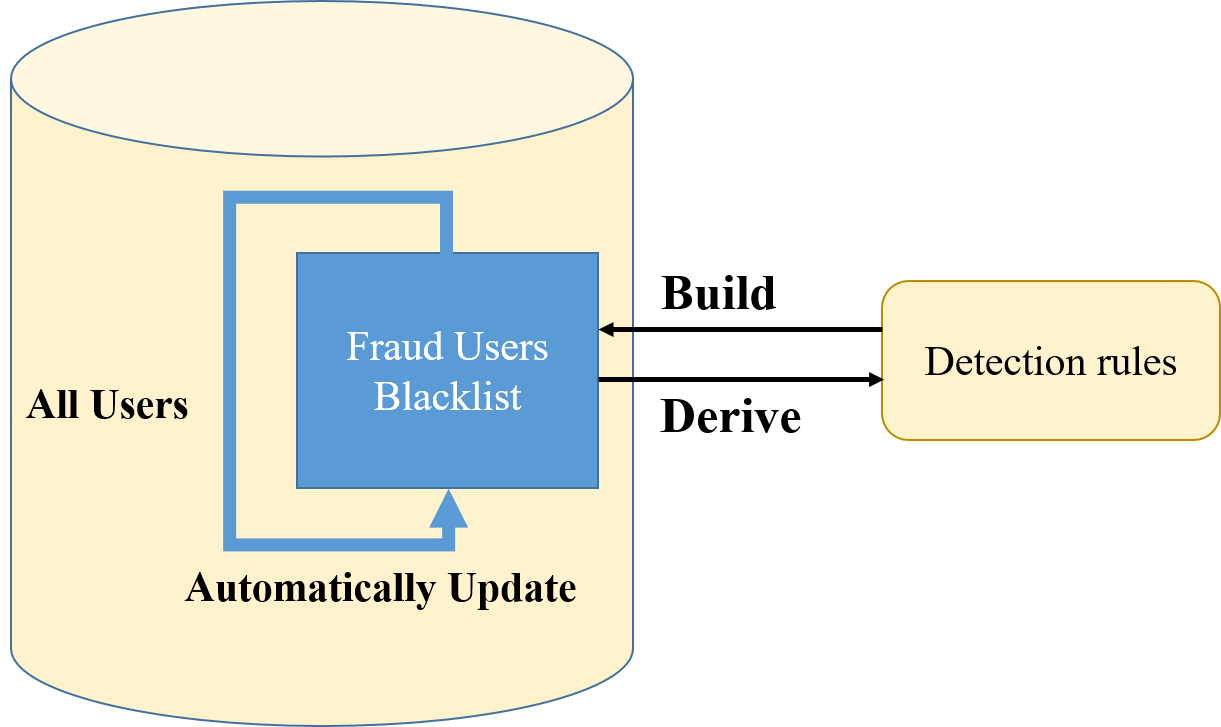}
			\label{figure:paradigm2}
		}
		\caption{Fraud detection paradigms}
		\label{figure:paradigm}
	\end{figure}

	Many digital payment platforms have been devoted to fraud detection for many years. These platforms have their own fraud users blacklists, and they track and analyze fraud users on the blacklists continuously. They have concluded many detection rules based on years of experience. As shown in Fig~\ref{figure:paradigm1}, platforms can use these detection rules to detect new fraud users and build a larger fraud user blacklist. It is impossible to gather all kinds of fraud users in this blacklist. There always exists unknown fraud users. These unknown fraud users cannot be detected based on existing detection rules. Moreover, if fraud users change their behavior patterns, they can escape from being detected by the platforms easily. In the traditional fraud detection paradigm, platforms cannot find new fraud patterns until new patterns of fraud users appear after they cause visible accidents on platforms.
	
	The limitations of traditional fraud detection paradigm are obvious. Platforms can only detect fraud users by existing knowledge. It is a passive defending paradigm. Platforms cannot update their detection rules to defense unknown fraud users automatically, which makes themselves vulnerable when new patterns of fraudsters appear. 
	
	\subsection{Improved Fraud Detection Paradigm}
	Traditional fraud detection paradigm cannot update detection rules until visible accidents happen, which will cause huge losses to platforms and other users. It is better to have an improved paradigm that can update its blacklist automatically. Platforms can derive new detection rules from new detected users. The improved fraud detection paradigm is shown in Fig~\ref{figure:paradigm2}.
	
	\begin{figure*}[ht]       
		\centering
		\includegraphics[width=\linewidth]{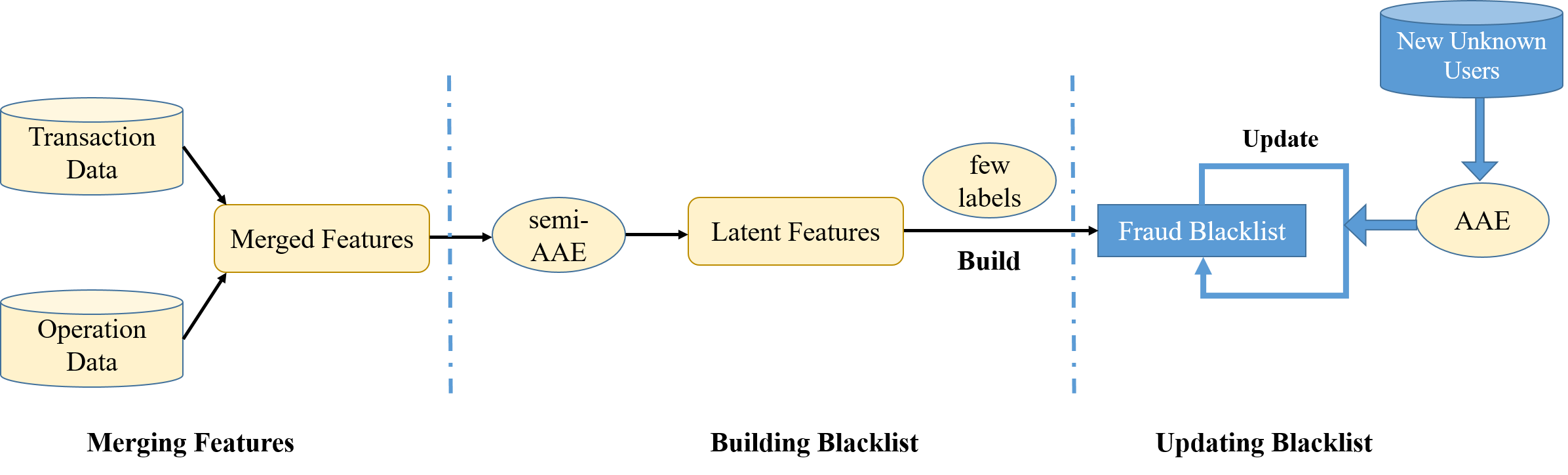}
		\caption{The overview of FraudJudger}  
		\label{figure:overview}
	\end{figure*}

	The most critical part of the improved paradigm is automatically updating fraud users blacklist. It requires the paradigm can detect new fraud users from unknown users, especially potential fraud users who can not be detected by existing detecting rules. Compared with traditional fraud detection paradigm, our improved paradigm can find new patterns of fraud users before they appear in large numbers and cause huge losses to others. The potential fraud users detection is offline, and the cost for the automatically updating part will not have much influence in practice.
	
	Based on the improved fraud detection paradigm, we design a creative fraud users detection model, FraudJudger. It can detect fraud users and actively identify fraud users beyond existing detecting rules from unknown users. 
	
	\section{FraudJudger: Fraud Detection Model}
	\label{section:Detection Model}

	\subsection{Model Overview}
	In this section, we introduce our fraud detection model, FraudJudger. As shown in the new detection paradigm, our model should contain two main functions: building fraud users blacklist and automatically updating the blacklist. The overview of our detection model is shown in Fig~\ref{figure:overview}. The whole detection model contains three phases:
	
	\textbf{Phase \uppercase\expandafter{\romannumeral1}:} Merging features from raw data. Platforms collect two main classes of information: operation data and transaction data. We first merge the two classes data, and the merged features are passed to our fraud detection part.
	
	\textbf{Phase \uppercase\expandafter{\romannumeral2}:} Building fraud users blacklist. Merged features in Phase \uppercase\expandafter{\romannumeral1} are of high dimensions, which can not be used directly. We can omit ineffective and noisy features and get efficient low dimension features with the help of adversarial autoencoder. In the meantime, we can detect fraud users by a few labeled data. It is a semi-supervised learning process. Detected fraud users will be added in the blacklist.
	
	\textbf{Phase \uppercase\expandafter{\romannumeral3}:} Updating fraud users blacklist by cluster analysis. The key idea is to find new fraud users beyond existing detection rules. In this phase, we train a new AAE network without labels to learn the latent representations of users. Then we cluster these latent variables from the network. After clustering latent variables, different users groups are formed. We detect fraud users with new patterns from these fraud groups.
	
	\subsection{Merging Features}
	
    Many electronic payment platforms record users' operation and transaction information. Operation data contain users' operation actions on payment platforms, such as device information, operation type  (changing password, viewing balance, etc), operation time, etc. Transaction data contain user's transaction information, such as transaction time, transaction amount, transaction receiver, etc.

	A user may leave many operation and transaction records on the electronic payment platform. We proposed an appropriate method to merge a user's records on the platform.
	
	We merge the two kinds of features by the key feature, which is "user id". It means that features belong to the same user will be merged. The value of each feature can be divided into two types, numeric and non-numeric. Numeric features can be analyzed directly. For non-numeric features, such as location information or device types, we use one-hot encoding method to convert them into a numeric vector. Each non-numeric feature is mapped into a discrete vector after one-hot encoding. If two features $f_{i}$ and $f_{j}$ have connections, we will construct a new feature $f_{new}$ to combine the two features. The new feature $f_{new}$ contains statistic properties of $f_{i}$ and $f_{j}$. 
	
	\subsection{Building fraud users blacklist}
	\label{subsection:builidng}
	In this phase, our model detects fraud users and build a fraud users blacklist. The main purposes of our models in this phase are:
	
	\begin{enumerate}
		\item Learning latent vectors of users.
		\item Detecting fraud users by latent vectors with a small scale of labels.
	\end{enumerate}
	
	The dimension of merged features from Phase \uppercase\expandafter{\romannumeral1} is too high to analyze directly for the following reasons:
	\begin{enumerate}
		\item Raw data contain irrelevant information, which is noise in our perspective. These irrelevant features will waste computation resources and affect the model's performance.
		\item High dimension features will weaken the model's generalization ability. Detection model will be easily overfitted. 
	\end{enumerate} 
	
	We should reduce the dimension of features and only leave essential features. Furthermore, if we manually choose important features based on experience, it will be hard to find new fraud patterns when fraud users change their attack patterns. 
	
\begin{figure}[h]       
		\centering
		\includegraphics[width=0.5\textwidth]{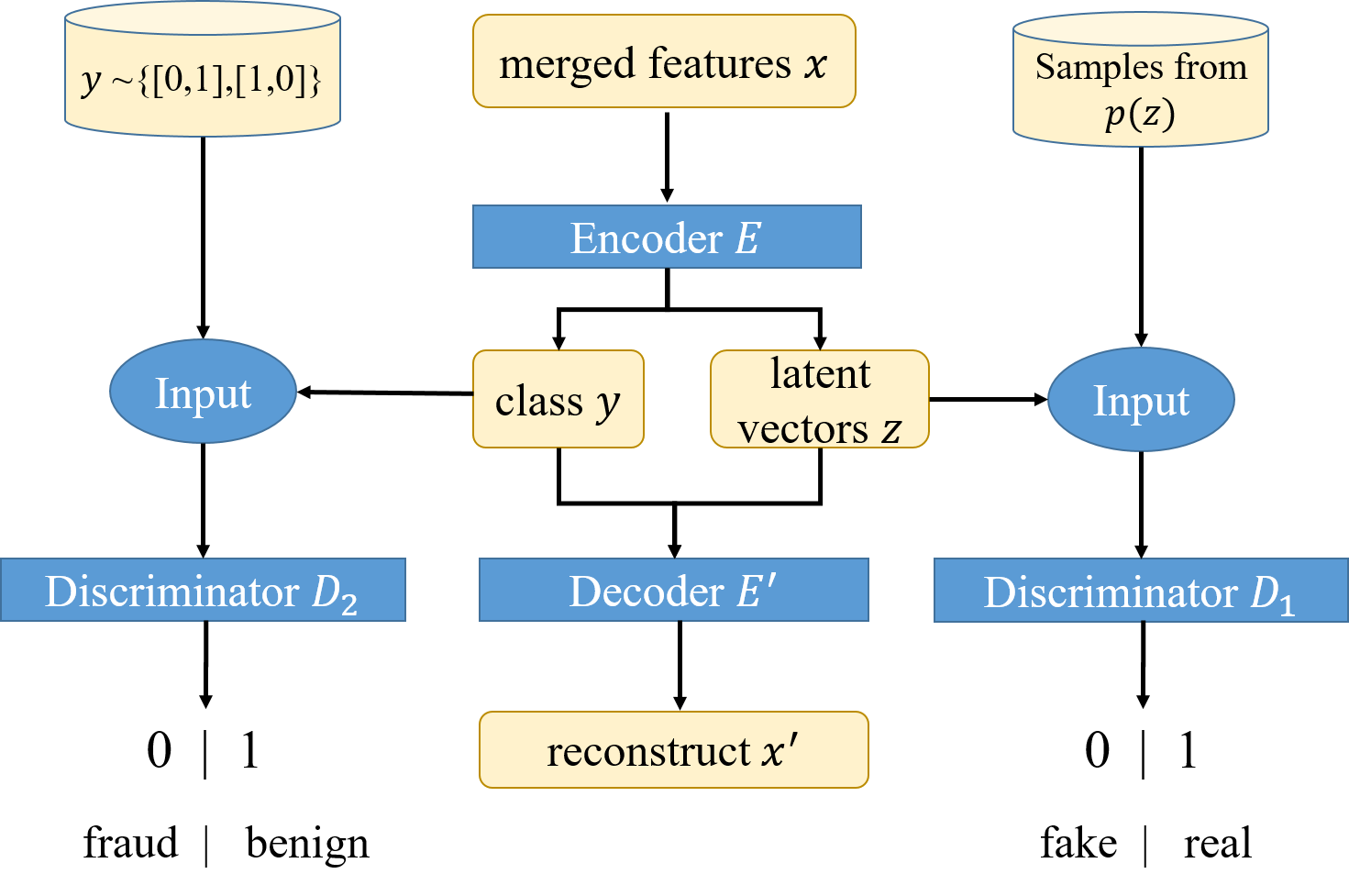}
		\caption{Architecture of semi-AAE network for fraud detection in FraudJudger}  
		\label{figure:semiaae}
	\end{figure}

	We build a semi-supervised AAE (semi-AAE) based fraud detection model to learn the latent representations of merged features and classify users. Our detection network contains three parts: encoder $E$, decoder $E'$ and two discriminators $D_1$ and $D_2$. Fig~\ref{figure:semiaae} shows the architecture of our detection network. 
	
    \textbf{Encoder:} For an input merged feature $x$, encoder $E$ will learn the latent representation $z$ of $x$. The dimension of the latent variables $z$ is less than the dimension of the input $x$, and the encoder's network structure determines it. The encoding procedure can be regarded as dimensionality reduction. Besides, it will output an extra one-hot variable $y$ to indicate the class of input value, which is a benign user or fraud user in our model. Our model uses $y$ to classify an unknown user. The inner structure of the encoder is a multi-layer network in our model.
	\begin{equation}
	E(x) = (y,z)
	\end{equation}
	
	\textbf{Decoder:} The purpose of the decoder is learning how to reconstruct the input of the encoder from encoder's outputs. The decoder's procedure is the inverse of the encoder. Inputs of the decoder $E'$ are outputs of the encoder $E$. The decoder will learn how to reconstruct inputs $x$ from $y$ and $z$. The output of the decoder is $x'$. The inner structure of the decoder is also the inverse of the inner structure of the encoder.
	\begin{equation}
	E'(y,z) = x'
	\end{equation}
	
	\textbf{Loss of Encoder-Decoder:} The loss of the encoder and the decoder $L_{e-d}$ is defined by mean-square loss between the input $x$ of the encoder and output $x'$ of the decoder. It measures the similarity between $x$ and $x'$.
	\begin{equation}
	L_{e-d} =  \mathbb{E} ((x-x')^2)
	\end{equation}
		
	\textbf{Generator:} Encoding the class $y$ and latent vectors $z$ from $x$ can be regarded as the generator. Let $p(y)$ be the prior distribution of $y$, which is the distributions of fraud users and benign users in the real world. And $p(z)$ is the prior distribution of $z$, which is assumed as Gaussian distribution:
	$z \sim  \mathcal{N}(\mu,\,\sigma^{2})$. The generator tries to generate $y$ and $z$ in their prior distributions to fool the discriminators. The loss function of the generator $L_G$ is:
	\begin{equation}
	L_{G} = -\mathbb{E}(log(1-D_1(z))+log(1-D_2(y)))\\
	\end{equation}
	
	\textbf{Discriminator:} Like the discriminator of GAN, we use discriminators in our model to judge whether a variable is real or fake. Since the encoder has two outputs, $y$ and $z$, we have two discriminators to discriminate them separately. The discriminators will judge whether a variable is in the real distribution. The loss function of discriminators $L_D$ are defined as:
	\begin{equation}
	\begin{split}
	L_{D_1} &= -\mathbb{E}(a_zlog(D_1(z))+(1-a_z)log(1-D_1(z)))\\
	L_{D_2} &= -\mathbb{E}(a_ylog(D_2(y))+(1-a_y)log(1-D_2(y)))\\
	L_D & = L_{D_1} + L_{D_2}
	\end{split}
	\end{equation}
	
	where $a_z$, $a_y$ are the true labels (fake samples or real) of inputs $z$ and $y$. The total loss of the discriminator part is the sum of each discriminator.
	
	\textbf{Classifier:} We can teach the encoder to output the right label $y$ with the help of a few samples with labels. And the loss function $L_C$ is:
	\begin{equation}
	L_C = -\mathbb{E}(a'_ylog(y)+(1-a'_y)log(1-y))
	\end{equation}
	
	where $a'_y$ means the right label (fraud samples or benign) for a sample, and $y$ is the output label from the encoder. When the encoder outputs a wrong label, the classifier will back-propagate the classification loss and teach the encoder how to predict right labels correctly.
	
	\textbf{Training Procedure:} The generator generates like the real label information $y$ and latent representations $z$ by the encoder network. Two discriminators try to judge whether the inputs are fake or real. It is a two-player min-max game. The generator tries to generate true values to fool discriminators, and discriminators are improving discrimination accuracy. Both of the generator and discriminators will improve their abilities simultaneously. For samples with labels, they can help to increase the classification ability of our model.
	
	Once the training of the semi-AAE model finishes, we can use it to classify users and build a fraud users blacklist.

	\subsection{Updating fraud users blacklist}
\begin{figure}[h]       
		\centering
		\includegraphics[width=0.5\textwidth]{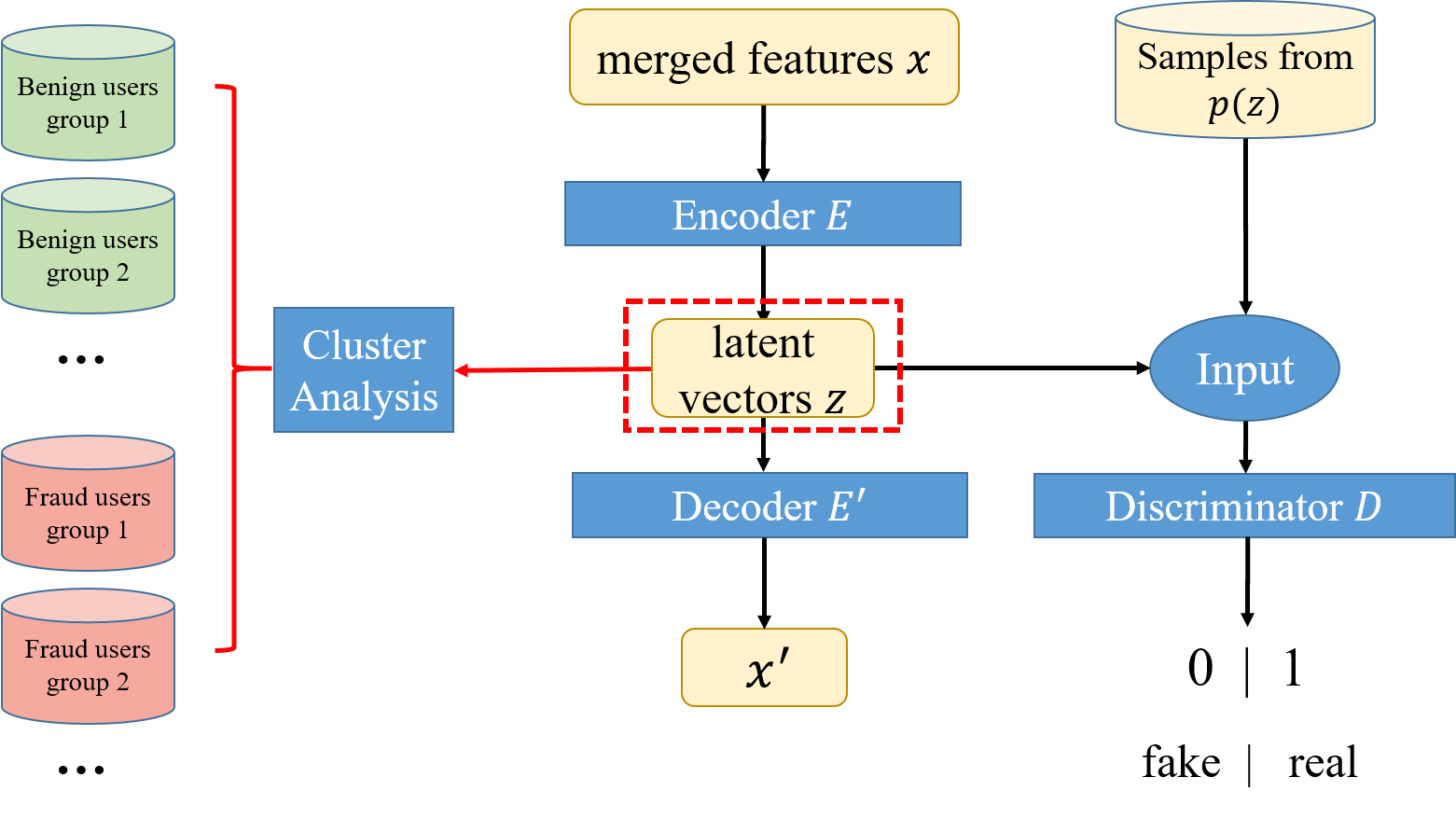}
		\caption{Architecture of detecting new fraud patterns in FraudJudger}  
		\label{figure:aae}
	\end{figure}

    The fraud detection part of FraudJudger can help us identify users based on rules we have known. We also hope that our model can detect users beyond existing detecting rules that we have known. In this section, we will teach our model how to learn unknown rules. The intuition is that we can identify potential fraud users with new fraud patterns. The architecture of the fraud updating part is shown in Fig~\ref{figure:aae}. It contains two parts, learning latent representations of unknown users by AAE network, and finding new fraud patterns from latent representations.
	
	\subsubsection{Learning latent representations from unknown users} First, we build another adversarial autoencoder network to learn latent representations of new users without labels. As we can see from Fig~\ref{figure:aae}, the network of learning latent representations in this phase has slight changes comparing with the network in Fig~\ref{figure:semiaae}.

	Since we need to find new fraud patterns from unknown users, we do not have label information in training data. We want to learn appropriate latent representation $z$ of input $x$ so that we can analyze users by $z$. The first change of the network is that the encoder only outputs latent representation $z$ instead of label information $y$. All information of a user is contained in the latent representation. The discriminator for label information is deleted as well. Since we have no data with labels, we do not have loss function for classification as well. The training phase of this network is also a two-player min-max game. 
	
	%    \subsubsection{Cluster analysis} Latent representations learned form AAE can well reflect a user's behavior pattern. We use cluster analysis on latent representations to identify each user's category, which means he is a benign user or not. The intuition for using cluster analysis is that fraud users' behavior patterns are different from benign users. Ideally, users can be divided into two groups by clustering. One is benign users group and another one for fraud users. However, it does not work in reality. Both of benign users and fraud users contain various kinds of behavior patterns on the electronic payment platforms. It is not reasonable to only use two cluster numbers. An appropriate clusters' number $k$ should be chosen. And users will be partitioned into $k$ groups. Each group contains users in a similar behavior pattern. In each group, we can give classification to each user by the semi-AAE classifier. Each group will be a benign user group or fraud user group depending on the subject category. If we find a user who is judged as a benign user in a fraud user group, he has a high probability of being a fraud user in fact. The procedure is shown in Algorithm~\ref{alg:potential}. 
	
	\subsubsection{Cluster analysis to find potential fraud users} 
	\label{subsubsection:cluster}
	After learning users' latent representations, we can find potential fraud users from unknown users by cluster analysis. Algorithm~\ref{alg:potential} shows the procedure of finding potential fraud users. 
	
	First, we classify unknown users by the classifier we construct in Section ~\ref{subsection:builidng}. Detected fraud users in this step are identified based on existing detection rules. Our aim is finding potential fraud users who have unknown fraud patterns. Potential fraud users will be wrongly classified as benign users in this step. We cluster users by their latent representations. Fraud users' behavior patterns are different from benign users'. Cluster analysis can cluster users with similar behavior patterns in the same group. We choose k-means as our cluster analysis method. Users will form $k$ clusters by k-means. Ideally, users can be divided into two groups by clustering, one is fraud users group, and the other one is benign users group. However, it does not work in reality. Both of benign users and fraud users contain various kinds of behavior patterns on the electronic payment platforms. It is not reasonable to only cluster users into two groups. An appropriate cluster number $n_{cluster}$ should be chosen. And users will be partitioned into $n_{cluster}$ groups. Each group contains users in a similar behavior pattern. Each group will be a benign user group or fraud user group depending on the ratio of detected fraud users. If the ratio is larger than a threshold $t_{fraud}$, we regard this group as fraud users group. If we find a user who is judged as a benign user in a fraud user group, he has high similarities with fraud users in his group, which indicates that he is more likely to be a fraud user in practice. 
	
	\begin{algorithm}[htb]
		\caption{Finding Potential Fraud Users}
		\label{alg:potential}
		\SetAlgoNoLine
		\KwIn{
			Set of unknown users $\mathbf{U} = \{u_1,u_2,..., u_n\}$  \;
			Classifier $Clf$ // The classification model  we constructed \;
			Cluster numbers $n_{cluster}$, the threshold of fraud users ratio $t_{fraud}$;}
		
		\KwOut{Potential fraud users set $\mathbf{U_p}$}
		
		\BlankLine
		Set of users' predicted labels $\mathbf{Y} = \{y_1,y_2,..., y_n\}$ \;
		Set of clustering group  $\mathbf{G} = \{g_1,g_2,..., g_{n_{cluster}}\}$ \;
		
		\For{$i=1,...,n$}
		{
			$y_i = Clf(u_i)$  // Classify each user by FraudJudger\;
		}
		$\mathbf{G} = Cluster(\mathbf{U},n_{cluster})$ // Cluster users to $n_{cluster}$ groups\;
		\For{$i=1,...,n_{cluster}$}
		{
			Calculate the ratio of fraud users $r_{faud}$\;
			\If{$r_{faud} > t_{fraud}$}
			{
				\For{user $u_j$ in group $i$}
				{
					\If{$y_j = $ benign users}
					{Add $u_j$ to $\mathbf{U_p}$}
				}        
			}
		}
		
	\end{algorithm}
	\DecMargin{1em}
	
	The intuition of our potential fraud users detection algorithm is that fraud users will gather together by clustering, and the learned latent representations ensure it. In traditional detection methods, features are manually chosen based on historical knowledge. We cannot find new fraud patterns in this way. In our FraudJudger model, we can learn latent representations of users without prior fraud detection knowledge. Even if fraud users change part of their fraud patterns, our model can still detect them.
	
	If potential fraud users are found, we can update our backlist and new detection rules can be derived by platforms.

	\section{Experiment}
	\label{section:Experiment}
	In this section, we first describe the real-world dataset we use in Section~\ref{subsection:data}. Then we evaluate FraudJudger's detection performance in Section~\ref{subsection:detectionperformance}. In order to intuitively showing the latent representations, we visualize the latent vectors of users in Section~\ref{subsection:visualization}. Next, we evaluate the performance of cluster analysis in Section~\ref{subsection:cluster}. Finally, we find the best dimension of latent representations in Section~\ref{subsection:dimension}.
	
	\subsection{Dataset Description}
	\label{subsection:data}
	We use a real-world dataset from Bestpay, which is a popular digital payment platform. The dataset has been anonymized before we use in case of privacy leakage. The dataset contains more than 29,000 user's operation behaviors and transaction behaviors in 30 days. All users in the dataset are manually labeled. We regard labels in this dataset as ground truth. In this dataset, the amount of fraud users is 4,046, which accounts for 13.78\% of total users. The dataset contains two kinds of data, one is operation data, and the other one is transaction data. There are 20 features in operation data and 27 features in transaction data. Some important operation features are listed in Table ~\ref{tbl:opr}, and part of important transaction features are listed in Table ~\ref{tbl:tra}.
	\begin{table}[htb]
		\centering
		\caption{Part features in operation data}
		\label{tbl:opr}
		\begin{tabular}{ccc}
			\toprule
			Feature  & Explanation               & Missing rate \\
			\midrule
			mode     & user's operation type     & 0\%           \\
			time     & operation time            & 0\%           \\
			device   & operation device          & 29.3\%        \\
			version  & operation version         & 19\%          \\
			IP       & device's IP address       & 18.0\%        \\
			MAC      & device's MAC address      & 89.9\%        \\
			os       & device's operation system & 0\%           \\
			geo\_code & location information      & 33.9\%        \\
			\bottomrule      
		\end{tabular}
	\end{table}
	
	\begin{table}[htb]
		\centering
		\caption{Part features in transaction data}
		\label{tbl:tra}
		\begin{tabular}{ccc}
			\toprule
			Feature  & Explanation               & Missing rate \\
			\midrule
			time     & transaction time            & 0\%           \\
			device   & transaction device          & 34.2\%        \\
			tran\_amt & transaction amount          & 0\%           \\
			IP       & device's IP address       & 14.6\%        \\
			channel  & platform type             & 0\%           \\
			acc\_id   & account id                & 62.1\%        \\
			balance  & balance after transaction & 0\%    \\
			trains\_type & type of transaction      & 0\% \\
			\bottomrule      
		\end{tabular}
	\end{table}
	
	As shown in Table ~\ref{tbl:opr} and Table ~\ref{tbl:tra}, there are some common features in both operation data and transaction data. We have some essential and useful information, like IP address, time, amounts, etc, to analyze a user's behavior. After merging features, we get 2174 dimensions of features for each user. Some features have a high missing rate, like $acc\_id$. We filter out features with a missing rate more than 30\%, and get 940-dimensional merged features for each user. FraudJudger will analyze the 940-dimensional merged features to detect frauds.
	
	\subsection{Detection Performance}
	\label{subsection:detectionperformance}
    In this experiment, we evaluate the detection ability of FraudJudger. First, we compare the model's performance with different proportions of labeled data. Then, we compare FraudJudger with other well-known supervised detection methods.
	
	\subsubsection{Different proportions of labeled data}
	
	We use 20,000 samples for training and the rest 9,354 samples for evaluating. To evaluate the performance of our model, we set five groups of experiments with different proportions of labeled samples:
	\begin{itemize}
		\item 1\% samples with labels.
		\item 2.5\% samples with labels.
		\item 5\% samples with labels.
		\item 10\% samples with labels.
		\item 25\% samples with labels.
	\end{itemize}
	
	The dimension of latent representations in our experiment is 100. We use a five-layer neural network as the structure of our encoder and decoder. The number of neurons in each hidden layer is 1024. Models are trained in 500 epochs, and the batch size in training is 200.
	
	We use four acknowledged standard performance measures to evaluate our model, which is precision, accuracy, recall and F-1 score. Precision is the fraction of true detected fraud users among all users classified as fraud users. Accuracy is the proportion of users who are correctly classified. Recall is intuitively the ability of the model to find all the fraud samples. F1-score is a weighted harmonic mean of precision and recall. 
	\begin{equation}
	\frac{1}{F1-score} = \frac{1}{Precsion} + \frac{1}{Recall}
	\end{equation}
	
\begin{figure}[h]       
		\centering
		\includegraphics[width=0.5\textwidth]{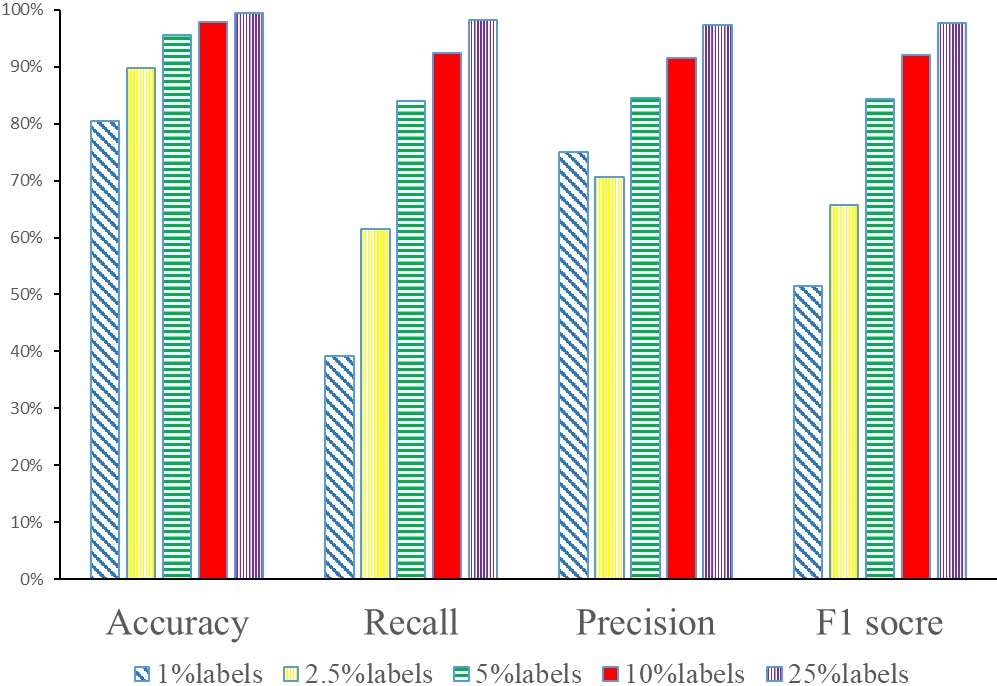}
		\caption{Accuracy, Precision, Recall and F1 Score of models}  
		\label{figure:res}
	\end{figure}

	And the result is shown in Fig~\ref{figure:res}. 
	
	Fig~\ref{figure:res} illustrates our model FraudJudger can achieve better performance with more labeled training samples. When the ratio of labeled samples is less than 10\%, the performance of the model increases rapidly as the ratio of labeled sample increases. When the ratio of labeled samples is more than 10\%, the increasing speed slows down. As we mentioned before, it is hard to obtain enough labeled samples to train our model in practical application. We do not need to use a large number of manually labeled data. Our experiment result shows that FraudJudger can achieve excellent classification performance with a small ratio of labeled data.
	
    \subsubsection{Compared with supervised models}

\begin{figure}[h]       
		\centering
		\includegraphics[width=0.5\textwidth]{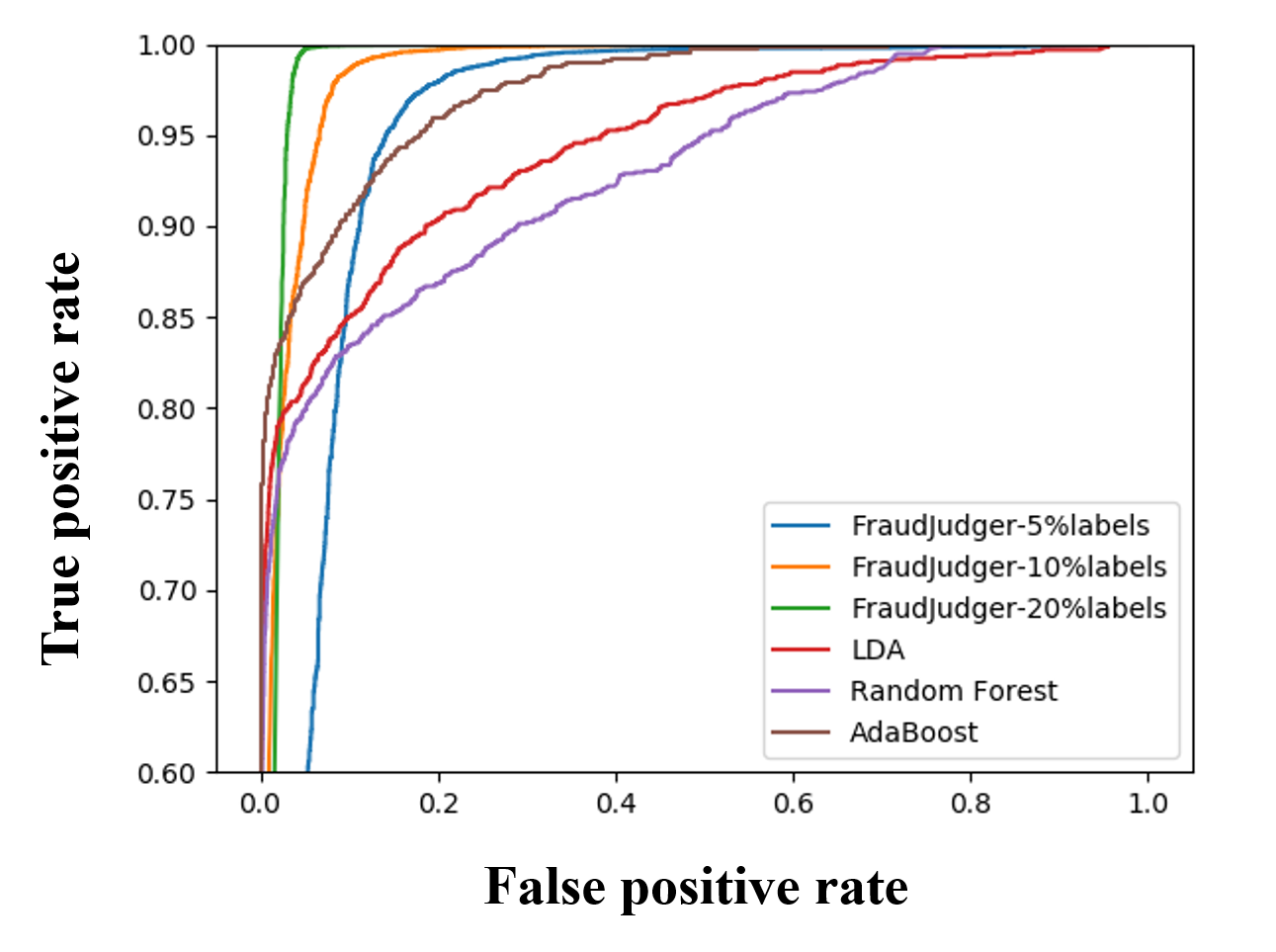}
		\caption{ROC of FraudJudger and other models}  
		\label{figure:roc}
	\end{figure}

	We compare our model's classification performance with other supervised classification models. Three different excellent machine learning models are chosen: Linear Discriminant Analysis (LDA), Random Forest and Adaptive Boosting model (AdaBoost). We set three groups of FraudJudger models with 5\% labels, 10\% labels, and 20\% labels, respectively. We use the ROC curve to evaluate the result. The result is shown in Figure ~\ref{figure:roc}, and the AUC of each model is shown in Table ~\ref{tbl:auc}.
	
%	\begin{table}[htb]
%		\centering
%		\caption{AUC of FraudJudger and other models}
%		\label{tbl:auc}
%		\begin{tabular}{c|ccccc}
%			\toprule
%			Models:  & \tabincell{c}{FraudAAE\\5\%labels} &   \tabincell{c}{FraudAAE\\10\%labels} & LDA & \tabincell{c}{Random\\Forest} & AdaBoost               \\
%			\midrule
%			AUC: &0.944&\textbf{0.983}&0.946&0.932&0.976 \\
%			\bottomrule      
%		\end{tabular}
%	\end{table}

	\begin{table}[htb]
	\centering
	\caption{AUC of FraudJudger and other models}
	\label{tbl:auc}
	\begin{tabular}{cc}
		\toprule
		Models  & AUC\\
		\midrule
		FraudJudger-5\%labels & 0.944\\
		\textbf{FraudJudger-10\%labels} & \textbf{0.983}\\
		\textbf{FraudJudger-20\%labels} & \textbf{0.985}\\
		LDA & 0.946\\
		Random Forest & 0.930\\
		AdaBoost & 0.975\\
		\bottomrule      
	\end{tabular}
	\end{table}
	
    As we can see from the result, the model's detection accuracy increases with more labeled training data. When the proportion of labeled data is larger than 10\%, FraudJudger outperforms all other supervised classification models. If we use fewer labels, FraudJudger still has satisfying performance. Compared with other supervised algorithms, we save more than 90\% work on manually labeling data and we achieve better performance.

	In conclusion, FraudJudger has an excellent performance on fraud users detection even with a small ratio of labeled data. Comparing with other supervised fraud detection methods, FraudJudger has a low requirement for the amount of labeled data. Our model can be applied in more realistic situations.

	\subsection{Visualization of Latent Representation}
	\label{subsection:visualization}
\begin{figure}[h]       
		\centering
		\includegraphics[width=0.5\textwidth]{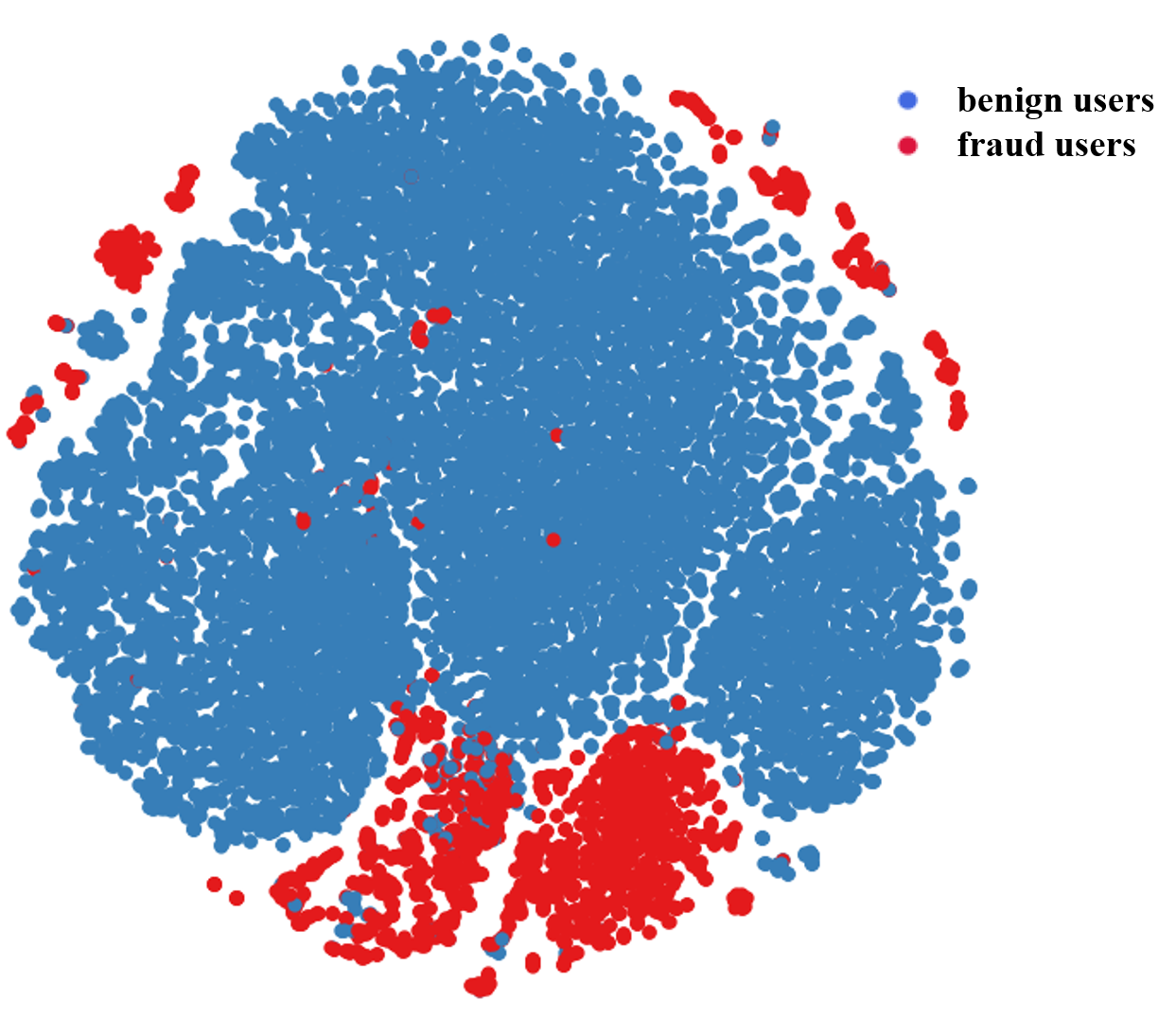}
		\caption{Visualization of latent representations by t-SNE}  
		\label{figure:tsne}
	\end{figure}

	FraudJudger uses learned latent representations to detect fraud users. In order to have an intuitively understanding of the latent representations, we use t-SNE \cite{maaten2008visualizing} to visualize the latent representations learned from FraudJudger. T-SNE is a practical method to visualize high-dimensional data by giving each data point a location in a two-dimensional map. Here we choose the dimension of latent representations equals to 100, and the ratio of labeled data is 10\%. The visualized result of t-SNE is shown in Fig~\ref{figure:tsne}.
	
	As we can see from Fig~\ref{figure:tsne}, the red points represent fraud users, and blue points represent benign users. Fraud users and benign users are well separated by latent representations. Benign users gather together and benign users are isolated to benign users. It means that the latent representations learned from FraudJudger can well separate benign users and fraud users.
	
\begin{figure}[h]       
		\centering
		\includegraphics[width=0.5\textwidth]{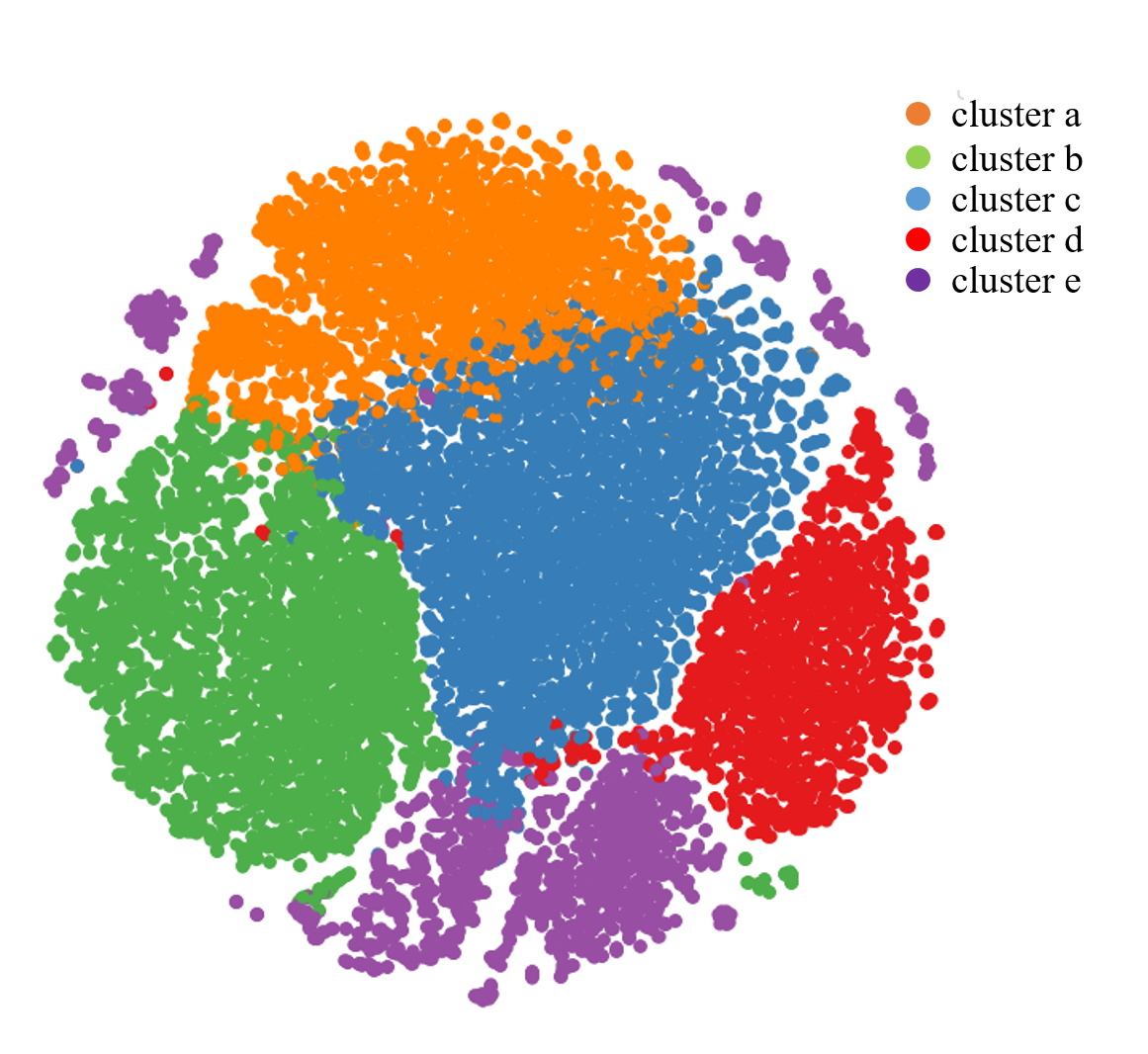}
		\caption{Visualization of cluster result of latent representations }  
		\label{figure:cluster}
	\end{figure}

	Furthermore, we cluster latent representations into five groups by K-means, and we visualize the result in Fig~\ref{figure:cluster}.
	
	Fig~\ref{figure:cluster} contains five different colors, and five colors representing five different groups of users. It is hoped that benign users and fraud users will form different groups after clustering, and the cluster result verifies it. The dividing lines between different groups are quite apparent. Comparing Fig~\ref{figure:cluster} with Fig~\ref{figure:tsne}, we can find that most fraud users are clustered into the same group in Fig~\ref{figure:cluster}. The fraud users in Fig~\ref{figure:tsne} are corresponding to the purple group in Fig~\ref{figure:cluster}. Benign users with different behavior patterns are clustered into four different groups. Fraud users and benign users are well separated by cluster analysis. 
	
	\subsection{Cluster Result}
	\label{subsection:cluster}
	In this section, we measure the performance of cluster analysis. As we mentioned in Section ~\ref{subsubsection:cluster}, we use cluster analysis to find potential fraud users. A more reasonable cluster result indicates that more likely to find potential fraud users. 
	
	After clustering, users are clustered into $n_{cluster}$ groups, and each group contains fraud users and benign users. If a group's fraud users ratio is larger than the threshold $t_{fraud}$, we regard this group as fraud group. However, different users may gather together because of other criterions, such as age, gender, etc. Different ages or genders of users will gather together instead of fraud users or benign users. We should measure whether users gather together with the right criterion.
	
	We propose a new measurement $\mathcal{R}$ called "\textit{Cluster Recall}" to measure the performance of the cluster result in gathering fraud users into the same group. $\mathcal{R}$  equals to the ratio of the number of fraud users in fraud groups and the number of all fraud users. A larger cluster recall indicates that more fraud users will gather into the same group, and the latent variables are better in representing fraud behaviors. A more reasonable learned latent representations will lead to a better cluster recall.
	
	\begin{equation}
	\mathcal{R} = \frac{the\, number\, of\,  fraud\, users \,in \,fraud \,groups}{the\, number\, of\, total\, fraud\, users}
	\end{equation}
	
	We set the threshold $t_{fraud}$ equals to 0.7. And we have four different models:\\
	\begin{enumerate}
		\item Origin features: merged features without dimensionality reduction
		\item DAE: latent representations learned by DAE \cite{vincent2010stacked} (Denoising Autoencoder)
		\item VAE: latent representations learned by VAE \cite{kingma2013auto} (Variational Autoencoder) 
		\item FraudJudger: our proposed model in learning latent representations
		%\item semi-AAE: our proposed semi-AAE based model
	\end{enumerate}
	
	In the first group, we cluster origin features directly. Both of DAE and VAE are well-known unsupervised methods to learn representations of a set of data. We use them as comparison models.  We use k-means as our cluster methods. We set $n_{cluster}=2,5,100$, and the result is shown in Table~\ref{tbl:clusterres}.
	
	\begin{table}[htb]
		\centering
		\caption{Cluster Recall for different models}
		\label{tbl:clusterres}
		\begin{tabular}{cccc}
			\toprule
			Number of cluster & 10    & 50   & 100  \\
			\midrule
			Origin Features    & 0.05    & 0.18 &0.36      \\
			DAE                & 0.14    & 0.43 & 0.55 \\
			VAE                & 0.09    & 0.32 & 0.42 \\
			\textbf{FraudJudger}       & \textbf{0.30}    & \textbf{0.48} & \textbf{0.59} \\
			%\textbf{semi-AAE } & 0.29 & 0.98 & 0.97 \\
			\bottomrule
		\end{tabular}
	\end{table}
	
	Table~\ref{tbl:clusterres} shows that our proposed model FraudJudger has a better performance than VAE and DAE in all three groups of different cluster numbers. It indicates that our model can better learn behavior patterns of fraud users. When the cluster number increases, the cluster recall is larger. However, it is meaningless if we use a too large number of clusters. If $n_{cluster}$ is too large, the number of users in each group after clustering will be small. It will be hard for digital payment platforms to analyze users' behavior patterns in this group. All of the dimensionality reduction models have a significant improvement compared with the group that uses origin features without dimensionality reduction. The experiment demonstrates that fraud users with similar fraud patterns will gather together in our model. We can use our model to find potential fraud users. If we set a  lower $t_{fraud}$, we can find more suspicious potential fraud users, and the credibility of suspicious potential users will drop. We should balance this in the real-world application.
	
	%When we set the threshold $t_{fraud}$ equals to 0.9, our model find 29 estimated potential fraud users from the real-world data by cluster analysis. If we set a lower $t_{fraud}$, we can find more suspicious potential fraud users, and the credibility of suspicious potential users will decrease. We should balance this in the real-world application.
	
	%When we set the threshold $t_{fraud}$ equals to 0.9, we find 74 suspicious potential fraud users from the real-world data by cluster analysis, and at least 29 of them are checked as fraud users by existing manually detection rules. It demonstrates that our model has the ability to find new fraud patterns. If we set a  lower $t_{fraud}$, we can find more suspicious potential fraud users, and the credibility of suspicious potential users will drop. We should balance this in the real-world application.
	
	\subsection{Dimension of Latent Representations}
	\label{subsection:dimension}
  	We analyze the performance of different dimensions of latent representations in FraudJudger to find the appropriate dimensions in practice. We use cluster recall $\mathcal{R}$ and adjust mutual information (AMI) \cite{vinh2010information} to measure models' performance in different dimensions.
	
	AMI is a variation of mutual information (MI) to compare two clusterings results. A higher AMI indicates the distributions of the two groups are more similar. For two groups A and B, the AMI is given as below:
	\begin{equation}
	AMI(A,B) = \frac{MI(A,B) - E(MI(A,B))}{avg(H(A), H(B)) - E(MI(A, B))}
	\end{equation}
	
	In our experiment, we calculate the AMI between data distribution of real data and data distribution of cluster results. We use 10\% labeled data when training FraudJudger. Ten groups of different dimensions of latent representations from 0 to 512 are conducted, and the result is shown in Fig ~\ref{figure:dimension}:
	\begin{figure}[h]       
		\centering
		\includegraphics[width=0.5\textwidth]{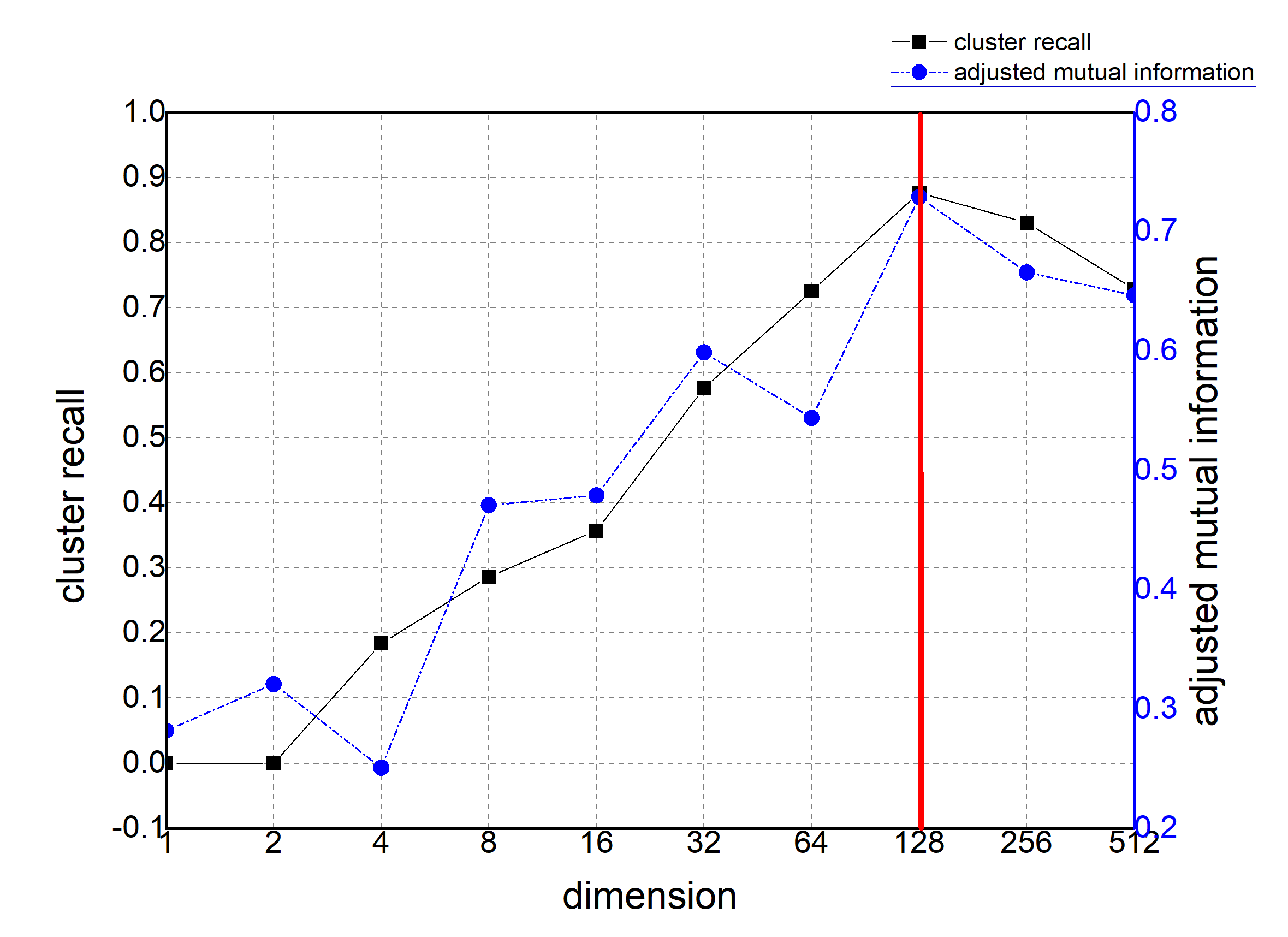}
		\caption{Performance of different dimensions of latent representations }  
		\label{figure:dimension}
	\end{figure}
	
    The result shows that the performance of FraudJudger varies with dimensions of latent representations. When the dimension is too low (dimension = 1 or 2), cluster recall is 0. Nearly no fraud users gather together in the same group. It shows that FraudJudger has poor performance in this case. Low dimensions of latent vectors cannot learn data distribution well. The model will lose many important features due to too low dimensions. When the dimension is higher, latent vectors contain more information about origin data, and the model's performance is better. When the dimension is around 128, FraudJudger has the best performance both in cluster recall and AMI. When the dimension of latent representations is too high, the performance decreases again. Under this circumstance, the model will learn a lot of noise features, which is harmful. Besides, a large dimension will lead to high model complexity and over-fitted. Thus, we should choose the dimension of latent representations around 128.
	
	\section{Conclusion}
	\label{section:Conclusion}
	In this paper, we proposed a novel fraud users detection model FraudJudger on real-world digital payment platforms. FraudJudger can learn latent features of users from original features and classify users based on the learned latent features. We overcome restrictions of real-world data, and only a few labeled training data are required. Fraud patterns are diverse and evolving, and our proposed method can be used in finding potential fraud users from unknown users, which is useful in anti-frauding. Our experiment is based on a real-world dataset, and the result demonstrates that FraudJudger has a good performance in fraud detection. Compared with other well-known methods, FraudJudger has advantages in learning latent representations of fraud users and saves more than 90\% manually labeling work. We have seen broad prospects of deep learning in fraud detection. 
	
	\bibliographystyle{ACM-Reference-Format}
	\bibliography{fraud}
	
\end{document}